\documentclass[12pt]{article}
\usepackage{amssymb,epsfig}

\renewcommand{\thefootnote}{\fnsymbol{footnote}}

\begin{document}

\title{
\begin{flushright}
\ \\*[-80pt] 
\begin{minipage}{0.2\linewidth}
\normalsize
hep-th/0512232 \\
KUNS-2004 \\*[75pt]
\end{minipage}
\end{flushright}
{\Large \bf 
Moduli-mixing racetrack model\\*[20pt]}}

\author{Hiroyuki~Abe\footnote{
E-mail address: abe@gauge.scphys.kyoto-u.ac.jp}, \ 
Tetsutaro~Higaki\footnote{
E-mail address: tetsu@gauge.scphys.kyoto-u.ac.jp} \ and \ 
Tatsuo~Kobayashi\footnote{
E-mail address: kobayash@gauge.scphys.kyoto-u.ac.jp} \\*[20pt]
{\it \normalsize 
Department of Physics, Kyoto University, 
Kyoto 606-8502, Japan} \\*[50pt]}

\date{
\centerline{\small \bf Abstract}
\begin{minipage}{0.9\linewidth}
\medskip 
\medskip 
\small
We study supersymmetric models with double gaugino 
condensations in the hidden sector, where the gauge 
couplings depend on two light moduli of superstring 
theory. We perform a detailed analysis of this class 
of model and show that there is no stable supersymmetric 
minimum with finite vacuum values of moduli fields. 
Instead, we find that the supersymmetry breaking occurs 
with moduli stabilized and negative vacuum energy. 
That yields moduli-dominated soft supersymmetry breaking terms.
To realize slightly positive (or vanishing) vacuum 
energy, we add uplifting potential. 
We discuss uplifting does not change qualitatively 
the vacuum expectation values of moduli and the above 
feature of supersymmetry breaking. 
\end{minipage}
}

\begin{titlepage}
\maketitle
\thispagestyle{empty}
\end{titlepage}


\renewcommand{\thefootnote}{\arabic{footnote}}
\setcounter{footnote}{0}

\section{Introduction}

String/M theory has been providing 
a lot of perspectives for particle physics.
However, it seems to have an infinite number 
of vacua and there are many moduli fields labelling them such 
as complex structure, volume of compact space, positions of 
branes, and so on. 
Vacuum expectation values (VEV) of these moduli fields 
determine various coupling constants and physical scales 
such as gauge couplings, Yukawa couplings and Planck scale. 
In order to find a realistic vacuum in string theory, it is 
necessary and important to determine phenomenologically 
reasonable values of them. 
Thus, the moduli stabilization is one of important 
issues to apply string theory to particle physics 
as well as cosmology. 

Besides the moduli stabilization problem, the origin of 
supersymmetry (SUSY) breaking is another puzzle which 
is in general related to the mechanism of stabilizing 
moduli fields.  It seems that some nontrivial mechanism 
is required in order to realize weak scale SUSY that 
can solve the gauge hierarchy problem, otherwise the 
SUSY breaking scale would be typically around the string scale 
or Planck scale. 
One of the elegant scenarios is to use 
field-theoretical dynamics like  
gaugino condensations~\cite{Affleck:1983mk} 
in the hidden sector, 
where the SUSY breaking scale can be suppressed by 
the dimensional transmutation of the gauge theory. 
A remarkable feature of the gaugino condensation 
scenario is that it can also generate a potential 
for moduli fields due to the fact that the gauge 
coupling is determined by their VEVs. 

The so-called racetrack model~\cite{Krasnikov:1987jj} is in 
most of the case based on the double gaugino condensations, 
and a modulus field can be stabilized thanks to the 
different modulus-dependences between two superpotential terms 
generated by them. However, in a simple setup, e.g. 
a single-modulus case, the resulting (local) minimum of the 
potential is a SUSY preserving anti-de Sitter (AdS) vacuum. 
Then, to be phenomenologically and cosmologically viable, 
we need at least two things here. One is SUSY breaking 
and the other is uplifting the vacuum energy to obtain 
a Minkowski or a slightly de Sitter (dS) background. 

Recently a way to achieve these two at the same time has 
been proposed in Ref.~\cite{Kachru:2003aw}, 
which is called the KKLT scenario. 
This scenario consists of two steps.
At the first step, a single modulus is stabilized 
because of the gaugino condensation and constant term 
in superpotential induced by three-form flux in type IIB 
string models, assuming that the other moduli are stabilized 
through flux compactification~\cite{Giddings:2001yu}. 
Such minimum corresponds to the SUSY AdS vacuum, 
as mentioned above.
At the second step, the uplifting of the vacuum energy is 
achieved by introducing anti $D3$-branes at the tip of a throat 
in the Calabi-Yau (CY) threefold which is highly warped 
due to the existence of three-form flux. 
Because the flux and anti $D3$-branes preserve 
different supercharges, SUSY is explicitly broken. 
This SUSY breaking effects are extensively studied 
in Ref.~\cite{Choi:2004sx}, and it has been shown 
that the resulting soft SUSY breaking terms of the visible fields 
have a quite distinctive pattern. 
(See for their phenomenological aspects Ref.~\cite{Choi:2005uz}.)
One of important points is that one can study analytically 
the AdS minimum of the potential before uplifting.
That makes it simple to understand the potential minimum 
with uplifting. 

The KKLT scenario is based on a quite simple setup of the 
type IIB orientifold models. A gauge kinetic function is a 
mixture of two or more moduli fields in several string 
models, e.g. weakly coupled heterotic string models~\cite{
Choi:1985bz,Ibanez:1986xy}, heterotic M models \cite{
Banks:1996ss,Choi:1997an,Nilles:1997vk,Buchbinder:2003pi}, 
type IIA intersecting D-brane models and type IIB 
magnetized D-brane models \cite{Cremades:2002te,
Lust:2004cx}.\footnote{
See also Ref.~\cite{Berg:2004ek} for 
moduli mixing among K\"ahler moduli, 
complex structure moduli and open string moduli 
in type IIB orientifold models.}
In this paper we consider general setup 
where the gauge couplings in the hidden sector 
are given by the mixture of some moduli fields 
(two moduli fields in practice).\footnote{ 
One of two moduli may be frozen around the string scale, 
e.g. by flux compactification.
Such scenario has been studied in Ref.~\cite{Abe:2005rx}.
Here we assume that both moduli remain light.}
On top of that, we consider the racetrack model, 
that double gaugino condensations generate 
nonperturbative superpotential terms and both of them 
depend on two moduli fields.
Based on the effective four-dimensional (4D) 
$N=1$ supergravity (SUGRA) description of such systems, 
we perform a detailed analysis of the moduli potential 
and investigate the structure of SUSY breaking and the moduli 
vacuum values at the local minimum. 
Similar models have been studied in the literature, 
in particular through numerical studies. 
We carry out detail study analytically under 
certain approximation. 
The potential minimum without adding an uplifting term
corresponds to SUSY breaking AdS vacuum.
Such detailed study on the potential minimum 
is as quite important as similar potential analysis 
at the first step of the KKLT scenario without 
uplifting is. 
Such study makes it possible to understand what would 
happen in our racetrack model after uplifting. 
Indeed, we discuss the potential minimum in our model with 
uplifting is qualitatively the same as one before uplifting. 
In our model, moduli-dominant SUSY breaking~\cite{Ibanez:1992hc,
Brignole:1993dj,Kobayashi:1994eh,Ibanez:1998rf} 
is realized, but the contribution due to anomaly 
mediation~\cite{Randall:1998uk} is negligible. 

The sections of this paper are organized as follows. 
In Sec.~\ref{sec:mixedgc}, we review several 
string models in which the gauge couplings are given 
by the mixture of moduli VEVs. 
Then assuming double gaugino condensations in the 
hidden sector, we analyze the racetrack model with 
such moduli-mixed gauge couplings within the framework 
of effective 4D $N=1$ SUGRA. 
First in Sec.~\ref{sec:effsugra}, we show the 
global structure of the moduli potential in such model. 
We study the stationary points of the potential in 
Sec.~\ref{sec:susy}, and find that the 
SUSY point is actually a saddle point. 
Then in Sec.~\ref{sec:susybreaking}, we show that 
there is a SUSY breaking local minimum close to the SUSY 
saddle point, and estimate the magnitude of SUSY breaking 
order parameters. 
In Sec.~\ref{sec:uplift}, 
we discuss the potential minimum after uplifting. 
In Sec.~\ref{sec:SUSYph}, we discuss SUSY breaking 
phenomenology. 
Sec.~\ref{sec:conclusion} is devoted to the conclusions 
and discussions. 
In Appendix~\ref{app:globalsusyvac}, we show that the SUSY point 
with $\langle W \rangle =0$ can be realized only at the 
runaway vacuum. 

\section{Moduli-mixed gauge couplings}
\label{sec:mixedgc}
In this paper, we consider general setup for 
the gaugino condensations where the gauge couplings 
are given by the mixture of some moduli fields.
Such situation occurs when we consider, e.g. 
heterotic models or type II models 
with intersecting/magnetized $D$-branes.
In these models, two moduli fields,  e.g. 
the dilaton $S$ and overall K\"ahler modulus $T$ 
appear in gauge kinetic function as their linear combination.
Here and hereafter, we call them moduli including the dilaton.
In this section, we review the heterotic 
and the magnetized $D$-brane cases in turn. 
Note that the proper definition of $S$ and $T$ depends on 
each string model.

In heterotic (M-)theory 
on\footnote{The symbol $\tilde{\times}$ represents
a simple direct product or including warp factor such as
$$ds^2_{10}=\Delta(y) g_{\mu \nu}dx^{\mu}dx^{\nu} 
+\Delta^{-1}(y)g_{mn}dy^m dy^n .$$} 
$M_4 \, \tilde{\times} \, CY_3 \, (\tilde{\times} S^1/Z_2)$, 
the one-loop gauge kinetic function of strong 
gauge group, $f_{strong}$, is given by~\cite{Buchbinder:2003pi}
\begin{eqnarray}
f_{strong}&=&S-\beta T +f_{M5}, 
\nonumber \\
\beta &\sim& 
\frac{1}{16\pi^4}\int_{CY}J\wedge 
\left[{\rm Tr}\,(F^{(2)})^2-\frac{1}{2}{\rm Tr}\,(R^2) \right], 
\nonumber
\end{eqnarray}
where $J$ is the K\"ahler form on CY with $h_{1,1}=1$. 
This can be seen from ten-dimensional Green-Schwarz term
$\int_{M_{10}} B_2 \wedge X_8$ 
or eleven-dimensional Chern-Simmons term 
$\int_{M_{11}} C_3 \wedge G_4 \wedge G_4$. 
The last term $f_{M5}$ represents the contribution from 
the M5-brane position moduli $Y \sim Tx^{11}_{M5}$ 
in the orbifold interval that is given by 
$f_{M5} = \alpha Y^2/T \sim \alpha T$ 
and $\alpha \sim \int_{CY}J \wedge *_6 J$. 
Hence, a gaugino condensation may generate 
a moduli-mixing superpotential, 
\begin{eqnarray}
W_{GC} \sim \exp[-(8\pi^2/N) f_{strong}]
\quad \textrm{ for } SU(N).
\nonumber
\end{eqnarray}

On the other hand, type II models such as intersecting 
D-brane models or magnetized D-brane models have  
gauge couplings~\cite{Lust:2004cx} similar to the above heterotic model. 
For example, in the supersymmetric type IIB magnetized D-brane 
model on $T^6/(Z_2 \times Z_2)$ orientifold\footnote{
Actually, the $T^6/(Z_2 \times Z_2)$ orbifold, 
in which the twist action is given by 
\begin{eqnarray}
\theta &:& (z^1,\, z^2,\, z^3) 
\ \rightarrow \ (-z^1,\, -z^2,\, z^3), 
\nonumber \\
\omega &:& (z^1,\, z^2,\, z^3) 
\ \rightarrow \ (z^1,\, -z^2,\, -z^3),
\nonumber
\end{eqnarray}
has three K\"ahler forms in the bulk, that is $h_{1,1}^{bulk}=3$. 
We here identify the indices of those cycles for simplicity.} 
with $h_{1,1}^{bulk}=1$, the gauge kinetic  functions are given as 
\begin{eqnarray}
f_{mD7} &=& |m_{7}|S+|w_{7}|T, 
\nonumber \\
f_{mD9} &=& m_9 S-w_9 T 
\quad \textrm{ for } O3/O7 \textrm{ system},
\nonumber
\end{eqnarray}
where the coefficients $m_p,\, w_p\, (p=7,9)\, \in {\mathbf Z}$ 
originate in Abelian magnetic flux contributions $F$ from the 
world volume and the Wess-Zumino term, and are given by 
$m_7 = \int_{mD7} F \wedge F$, 
$m_9 = \int_{mD9} F\ \wedge F \wedge F $ and 
$w_9 = \int_{mD9} *_6{J}_{bulk}\wedge F$ up to a numerical factor. 
The $w_{9}$ corresponds to the winding number on a wrapping 
4-cycle and magnetic flux contribution, while $w_7$ corresponds 
to the winding number of D7-brane on the 4-cycle. The signs of $m_9$ 
and $w_9$ depend on the magnetic fluxes and SUSY conditions. 
Notice that the Abelian gauge magnetic flux $F$ is quantized 
on a compact 2-cycle $C_2$ as $\int_{C_2}F \in {\mathbf Z}$ in 
this case. For example in Ref.~\cite{Marchesano:2004xz}, one can 
find negative $m_9$ and $w_9$. In addition, T-duality action 
can exchange winding number for magnetic number, but the result 
is similar, that is 
\begin{eqnarray}
f_{mD9}&=&W_9 S-M_9 T 
\qquad \textrm{ for }O9/O5 \textrm{ system}, 
\nonumber
\end{eqnarray} 
where $W_9,~M_9~\in {\mathbf Z}$ are, respectively, 
the winding number on the 6-cycle, and 
the winding number on the 2-cycle and 
magnetic flux contributions given by 
$M_{9}=\int_{mD9} {J}_{bulk} \wedge F \wedge F$. 
Again we have neglected numerical factors. 

The gauge coupling on the magnetized brane is written by 
\begin{eqnarray}
\frac{1}{g_{mD9}^2} &=& 
\big| m_9 {\rm Re}\,S- w_9 {\rm Re}\,T \big|. 
\nonumber
\end{eqnarray}
The magnetic fluxes can contribute to RR tadpole 
condition of four-form potential and eight-form 
potential~\cite{Marchesano:2004xz, Cascales:2003zp}. 
In this paper, we will treat $w_p$ and $m_p$ as free parameters. 

In type IIA intersecting D-brane models, which are T-duals 
of the above IIB string models, the above expressions of 
K\"ahler moduli change for complex structure moduli. 
However, since there are three- and even-form fluxes, 
all geometric moduli can be frozen in these type IIA models 
at low energy as a supersymmetric AdS vacuum. 

In orbifold string theory, moduli in twisted sectors, the 
so-called twisted moduli $M$ can exist~\cite{Aldazabal:1998mr}. 
These modes can contribute to the gauge kinetic function on 
D-branes near orbifold fixed points, 
\begin{eqnarray}
f_{Dp}= (S ~{\rm or}~T) + \sigma M, 
\nonumber
\end{eqnarray} 
where $\sigma$ is ${\cal O}(0.1)$-${\cal O}(1)$ parameter 
depending on gauge and orbifold group. 
These twisted moduli may be stabilized easily due to their 
K\"ahler potential~\cite{Abel:2000tf}, but make little contribution 
to the gauge coupling because the moduli are related to collapsed 
cycles of orbifold. Then we may naturally have $\langle M \rangle \ll 1$, 
and neglect contributions of those moduli. 

We comment that if a contribution of $T$ in the gauge coupling 
$f=xS \pm yT$ is required to be small compared with $S$, 
one needs to tune the ratio $y/x$ for a few percent. 
For example in heterotic M-theory, we need to choose a moderate 
CY model or to tune positions of M5-branes or the magnetic flux 
and winding number of D-branes. 

\section{Effective 4D $N=1$ SUGRA}
\label{sec:effsugra}
Motivated by the moduli-mixed gauge couplings explained in 
the previous section, we consider a racetrack model 
with double gaugino condensations at the hidden sector where 
the gauge couplings depend on two light moduli\footnote{
If, e.g. there exists three-form flux as in the KKLT model, 
one of moduli, say $S$, can be stabilized around the string scale. 
In this case the modulus field 
$S$ should be replaced by the vacuum value 
$\langle S \rangle$ in the effective theory, 
and the analysis is quite different from the 
one in this paper. 
Such scenario has been closely studied 
in Ref.~\cite{Abe:2005rx}.} 
represented by $S$ and $T$. 
We analyze a scalar potential of the effective 
4D $N=1$ SUGRA characterized by the K\"ahler 
and the superpotential, 
\begin{eqnarray}
K &=& -n_S \ln (S+\bar{S})-n_T \ln (T+\bar{T}), 
\nonumber \\
W &=& W_1+W_2, 
\nonumber
\end{eqnarray}
where $n_S$, $n_T>0$ and 
\begin{eqnarray}
W_1 &=& A e^{-a(S-\alpha T)}, 
\nonumber \\
W_2 &=& -B e^{-b(S+\beta T)}. 
\nonumber
\end{eqnarray} 
We parameterize the effective theory 
by $n_S$, $n_T$, $a$, $b$, $\alpha$, $\beta$, 
$A$ and $B$ which depend on the hidden gauge groups 
and the string models explained in the previous section. 
The scalar potential of this system is 
written in the standard $N=1$ SUGRA form as 
\begin{eqnarray}
V &=& e^G(G^{I\bar{J}}G_IG_{\bar{J}}-3) 
\ = \ K_{I\bar{J}}F^I \bar{F}^{\bar{J}}-3e^K|W|^2, 
\label{eq:scalarpotential}
\end{eqnarray}
where $I,J=(S,T)$, $G=K+\ln|W|^2$, 
$\bar{F}^{\bar{I}}=-e^{K/2} K^{\bar{I}J} D_J W$ and 
\begin{eqnarray}
D_S W &\equiv& G_S W \ = \ 
(b-a)Ae^{-a(S-\alpha T)}+(K_S-b)W, 
\nonumber \\
D_T W &\equiv& G_T W \ = \ 
(a\alpha+b\beta)Ae^{-a(S-\alpha T)}+(K_T-b\beta)W. 
\nonumber
\end{eqnarray}
The complex scalar fields $S$ and $T$ are written as 
$$S=s+i\sigma_s,\qquad T=t+i\sigma_t,$$
respectively by using four real scalars 
$s$, $t$, $\sigma_s$ and $\sigma_t$. 
The SUSY conditions $D_S W=D_T W=0$ can have a runaway 
solution with $|s|$ or $|t| \rightarrow \infty$ 
and $\langle W \rangle=0$. 
(See Appendix~\ref{app:globalsusyvac}.)
This case is outside our interests and we will find 
a nontrivial vacuum with $\langle W \rangle \ne 0$ 
in the following analysis.

\subsection{Imaginary directions}
First we show that $\sigma_s$ and $\sigma_t$ are decoupled 
from $s$ and $t$ by their stationary conditions. 
The scalar potential (\ref{eq:scalarpotential}) 
can be written as 
\begin{eqnarray}
V &=& e^K \bigg\{ 
\bigg( \frac{1}{K_{S\bar{S}}}(K_S-a)
+\frac{1}{K_{T\bar{T}}}(K_T+a \alpha)-3 \bigg)|W_1|^2 
\nonumber \\ 
&& \qquad 
+\bigg( \frac{1}{K_{S\bar{S}}}(K_S-b)
+\frac{1}{K_{T\bar{T}}}(K_T-b \beta)-3 \bigg)|W_2|^2 
\nonumber \\ 
&& \qquad 
+r(s,t)(\bar{W}_1W_2+\textrm{h.c.}) \bigg\}, 
\nonumber
\end{eqnarray}
where 
\begin{eqnarray}
r(s,t) &=& 
\frac{1}{K_{S\bar{S}}}(K_S-a)(K_S-b)
+\frac{1}{K_{T\bar{T}}}(K_T+a \alpha)(K_T-b \beta)-3, 
\nonumber
\end{eqnarray}
and we easily find that the following term  
$$\bar{W}_1W_2+\textrm{h.c.} \ = \ 
2\,{\rm sign}(AB)\,
|W_1||W_2|\cos \big[ (b-a)\sigma_s
+(a \alpha+b \beta)\sigma_t \big],$$ 
is only the source of the potential for 
$\sigma_s$ and $\sigma_t$. 
The stationary conditions 
$\partial_{\sigma_s}V=\partial_{\sigma_t}V=0$ fix 
a linear combination of $\sigma_s$ and $\sigma_t$, 
\begin{eqnarray}
(b-a)\sigma_s+(a \alpha+b \beta)\sigma_t &=& n\pi 
\qquad (n: \textrm{ integer}), 
\label{eq:imaginaryvev}
\end{eqnarray}
while another combination remains as a flat direction. 
Along this stationary direction, we also find 
$\partial_i \partial_{\sigma_j}V=0$ where $i,j=(s,t)$. 
That means that there is no mixing between 
$(s,t)$ and $(\sigma_s,\sigma_t)$. Then we can analyze 
the stability of $s$, $t$ separately 
from $\sigma_s$, $\sigma_t$ in the following sections. 
Note that the even or odd $n$ corresponds to the (local) 
minimum or maximum of the potential, 
depending on ${\rm sign}\big( AB\,r(s,t) \big)$, 
i.e. depending on the vacuum values of $s$ and $t$ shown later. 

The remaining flat direction is expected to receive 
typically a mass of ${\cal O}(m_{3/2}/4\pi)$ 
at the one-loop level with some SUSY breaking. 
We will not discuss about this flat direction 
in this paper, although it can play a role in 
phenomenological and cosmological arguments 
(see, e.g. Ref.~\cite{Barreiro:1999hp} and references therein).

\subsection{Real directions}
Next we show the global structure of the scalar potential 
$V(s,t)$ where $\sigma_s$ and $\sigma_t$ are already fixed 
by Eq.~(\ref{eq:imaginaryvev}). 
The $F^S$-flat direction, 
\begin{eqnarray}
D_S W
&=& (K_S-a)Ae^{-a(S-\alpha T)}-(K_S-b)Be^{-b(S+\beta T)}
\ = \ 0, 
\nonumber
\end{eqnarray}
is determined by the curve 
\begin{eqnarray}
t &=& 
\frac{1}{a\alpha+b\beta}
\bigg( \ln 
\frac{B(2b s+n_S)}{A(2a s+n_S)} 
-(b-a) s \bigg), 
\nonumber
\end{eqnarray}
with Eq.~(\ref{eq:imaginaryvev}), which has asymptotic lines 
\begin{eqnarray}
\left\{ 
\begin{array}{rcll}
t &\to& \displaystyle 
-\frac{b-a}{a\alpha+b\beta} s 
+\frac{1}{a\alpha+b\beta} \ln \frac{bB}{aA}, 
& (s \to \pm \infty,\ bB/aA>0), \\*[10pt]
s &\to& \displaystyle 
-\frac{n_S}{2a},\quad -\frac{n_S}{2b}, 
& (t \to \pm \infty). 
\end{array}
\right.
\label{eq:dswal}
\end{eqnarray}
Similarly, the $F^T$-flat direction, 
\begin{eqnarray}
D_T W &=& 
(K_T+a\alpha)Ae^{-a(S-\alpha T)}-(K_T-b\beta)Be^{-b(S+\beta T)} 
\ = \ 0, 
\nonumber
\end{eqnarray}
draws the curve
\begin{eqnarray}
s &=& 
\frac{1}{b-a}
\bigg( \ln 
\frac{B(-2b\beta t-n_T)}{A(2a\alpha t-n_T)} 
-(a\alpha+b\beta) t \bigg), 
\nonumber
\end{eqnarray}
with asymptotic lines 
\begin{eqnarray}
\left\{ 
\begin{array}{rcll}
t &\to& \displaystyle 
-\frac{n_T}{2a\alpha},\quad -\frac{n_T}{2b\beta}, 
& (s \to \pm \infty), \\*[10pt]
s &\to& \displaystyle 
-\frac{a\alpha+b\beta}{b-a} t 
+\frac{1}{b-a} \ln \frac{bB}{aA} 
+\frac{1}{b-a} \ln \frac{-\beta}{\alpha}, 
& (t \to \pm \infty,\ b\beta B/a\alpha A<0). 
\end{array}
\right.
\nonumber
\end{eqnarray}

A certain linear combination of 
$D_S W$ and $D_T W$ is found to be 
\begin{eqnarray}
(a\alpha+b\beta)D_S W-(b-a) D_T W &=& h(s,t)\,W, 
\nonumber
\end{eqnarray}
where 
\begin{eqnarray}
h(s,t) &\equiv& 
(a\alpha+b\beta)K_S+(a-b)K_T-ab(\alpha+\beta). 
\nonumber
\end{eqnarray}
The solution of $\langle W \rangle=0$ corresponds to 
only the runaway vacuum with $|s|$ or $|t| \rightarrow \infty$. 
(See Appendix~\ref{app:globalsusyvac}.) 
We are not interested in such a runaway solution 
$\langle W \rangle=0$, and then for $b \ne a$ 
and $a\alpha+b\beta \ne 0$, one of the SUSY 
conditions $D_S W=0$ and $D_T W=0$ can be 
replaced by $h(s,t)=0$ resulting 
\begin{eqnarray}
t &=& 
\frac{n_T (b-a) s}{2ab(\alpha+\beta)s
+n_S(a\alpha+b\beta)}. 
\nonumber
\end{eqnarray}
This curve always passes through the origin 
of $(s,t)$-plane, and has asymptotic lines 
\begin{eqnarray}
\left\{ 
\begin{array}{rcll}
t &\to& \displaystyle t_\infty \equiv 
\frac{n_T(b-a)}{2ab(\alpha+\beta)}, 
& (s \to \pm \infty), \\*[10pt]
s &\to& \displaystyle s_\infty \equiv 
-\frac{n_S(a\alpha+b\beta)}{2ab(\alpha+\beta)}, 
& (t \to \pm \infty). 
\end{array}
\right. 
\label{eq:dswdtwal}
\end{eqnarray}

For $\sigma_s$ and $\sigma_t$ satisfying 
Eq.~(\ref{eq:imaginaryvev}), we have relations, 
\begin{eqnarray}
\frac{W_2}{W_1} &=& 
-\frac{a}{b}e^{-\Phi_\bot(s,t)}, 
\quad 
\frac{\partial_S W_2}{\partial_S W_1} \ = \ 
-e^{-\Phi_\bot(s,t)}, 
\quad 
\frac{\partial_T W_2}{\partial_T W_1} \ = \ 
\frac{\beta}{\alpha} e^{-\Phi_\bot(s,t)}, 
\quad \ldots, 
\label{eq:wratio}
\end{eqnarray}
where 
\begin{eqnarray}
\Phi_\bot(s,t) &=& 
(b-a)s+(a\alpha+b\beta)t
-\ln \frac{bB}{aA}. 
\label{eq:massivemode}
\end{eqnarray}
Here the ellipsis denotes similar relations for higher derivatives.
That implies that our system is almost described 
by the superpotential $W \simeq W_1$ ($W \simeq W_2$) 
in the region $\Phi_\bot(s,t) \gg 1$ ($\Phi_\bot(s,t) \ll -1$) 
without large hierarchies between parameters, 
$$a/b,\ \alpha/\beta \ \sim \ {\cal O}(1).$$ 
In the band $-1 \lesssim \Phi_\bot(s,t) \lesssim 1$, 
the contributions from $W_1$ and $W_2$ are comparable, 
and actually this area is spreading along the 
asymptotic line of $D_S W=0$ shown in Eq.~(\ref{eq:dswal}), 
which corresponds to $\Phi_\bot(s,t)=0$. \\

\begin{figure}[t]
\begin{center}
\epsfig{figure=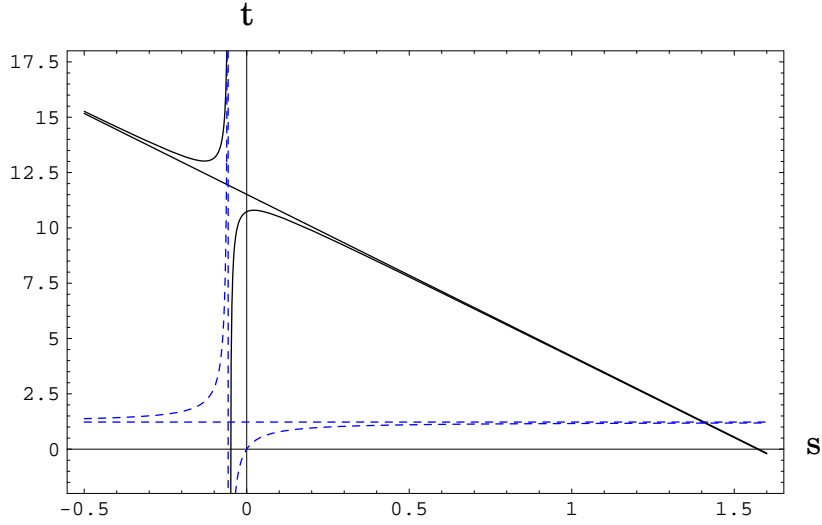,width=0.7\linewidth}
\end{center}
\caption{The curves of $D_SW=0$ (solid curve) and 
$h(s,t)=0$ (dotted curve) with each asymptotic lines 
in $(s,t)$-plane. 
The parameters are chosen as in Eq.~(\ref{eq:numpara}). 
The cross-point between the solid and dotted curve 
corresponds to the SUSY stationary point.}
\label{fig:flatdir}
\end{figure}

In Fig.~\ref{fig:flatdir}, we show the curves 
of $D_SW=0$ and $h(s,t)=0$ together with the asymptotic 
lines (\ref{eq:dswal}) and (\ref{eq:dswdtwal}) respectively 
in the $(s,t)$-plane with Eq.~(\ref{eq:imaginaryvev}), 
for a parameter choice, 
\begin{eqnarray}
a &=& \frac{8\pi^2}{N_1}, \quad 
b \ = \ \frac{8\pi^2}{N_2}, \quad 
\alpha \ = \ \frac{N_1}{8\pi^2}n_1, \quad 
\beta \ = \ \frac{N_2}{8\pi^2}n_2, 
\nonumber
\end{eqnarray}
and 
\begin{eqnarray}
N_1 &=&   9, \quad 
N_2 \ = \ 8, \quad 
n_1 \ = \ 1/40, \quad 
n_2 \ = \ 1/8, 
\nonumber \\
A   &=&   1.00 \times 10^{-6}, \quad 
B   \ = \ 5.00 \times 10^{-6}, \quad 
n_S \ = \ 1, \quad 
n_T \ = \ 3. 
\label{eq:numpara}
\end{eqnarray}

The asymptotic values (\ref{eq:dswdtwal}) tell us that, 
when both of $t_\infty$ and $s_\infty$ are negative, 
there is no SUSY point within the physical region 
$s,t \ge 0$ of the moduli space. 
In this case the scalar potential (\ref{eq:scalarpotential}) 
just has a runaway structure or is unbounded from below, 
without a nontrivial stationary point for $s,t \ge 0$. 
When $t_\infty > 0$ and/or $s_\infty > 0$, we have a possibility 
of SUSY stationary point in the region $s,t > 0$. 
Thus, in the following, we consider such case, in which the 
SUSY stationary point ($s_{SUSY}$,\,$t_{SUSY}$) can exist 
within the physical domain $s_{SUSY}$, $t_{SUSY}>0$. 
Moreover, in the region 
\begin{eqnarray} 
|s_{SUSY}| &\gg& \Big| \frac{n_S}{2a} \Big|,\ 
\Big| \frac{n_S(a\alpha+b\beta)}{2ab(\alpha+\beta)} \Big|, 
\nonumber
\end{eqnarray}
the SUSY point is approximately located at the cross-point of 
two asymptotic lines (\ref{eq:dswal}) and (\ref{eq:dswdtwal}), that is 
\begin{eqnarray}
s_{SUSY} 
&\simeq& \frac{1}{b-a}\ln\frac{bB}{aA}
-\frac{a\alpha+b\beta}{b-a}t_{SUSY}, 
\qquad 
t_{SUSY} 
\ \simeq \ \frac{n_T(b-a)}{2ab(\alpha+\beta)}. 
\nonumber
\end{eqnarray}
In addition, if we assume 
\begin{eqnarray}
|s_{SUSY}| &\gg& \Big| \frac{a\alpha+b\beta}{b-a} \Big| 
|t_{SUSY}|, 
\nonumber
\end{eqnarray}
we obtain a simple form 
\begin{eqnarray}
s_{SUSY} 
&\sim& \frac{1}{b-a}\ln\frac{bB}{aA}, 
\qquad 
t_{SUSY} 
\ \simeq \ \frac{n_T(b-a)}{2ab(\alpha+\beta)}. 
\label{eq:sspal}
\end{eqnarray}
The vacuum energy is negative at this point, 
\begin{eqnarray}
V_{SUSY} &=& -3(m_{3/2}^{SUSY})^2 \ < \ 0, 
\nonumber
\end{eqnarray}
where 
\begin{eqnarray}
m_{3/2}^{SUSY} &\sim& 
\frac{b-a}{b} 
\Big( \frac{aA}{bB} \Big)^{\frac{a}{b-a}}
\frac{A}{(2s_{SUSY})^{n_S/2} (2t_{SUSY})^{n_T/2}}. 
\label{eq:m32susyads}
\end{eqnarray}
Note that the gravitino mass is nonvanishing 
at the AdS SUSY point. 

In the following we consider the case that the 
parameters $A$, $B$, $a$, $b$, $\alpha$ and $\beta$ 
satisfy\footnote{For the parameters satisfying 
$s_{SUSY}<0$ and/or $t_{SUSY}<0$, the potential 
has trivial structure in the domain $s,t \ge 0$. 
In this case the moduli can not be stabilized at finite values.} 
$s_{SUSY}>0$ and $t_{SUSY}>0$ in Eq.~(\ref{eq:sspal}). 
For small values of $N_1$ and $N_2$, the parameters 
$a=8\pi^2/N_1$ and $b=8\pi^2/N_2$ are larger than unity, 
and then $s_{SUSY}$ and $t_{SUSY}$ tend to be smaller than 
unity in the unit $M_{Pl}=1$. Our tree level effective 
SUGRA analysis is not reliable in this region. 
For the parameters satisfying, e.g. 
\begin{eqnarray}
a,b \ \gg \ 1, \qquad 
b-a \ \sim \ {\cal O}(1), \qquad 
B/A \ > \ a/b, \qquad 
\alpha, \beta \ \sim \ {\cal O}(1/ab), 
\label{eq:pararef}
\end{eqnarray}
the approximation (\ref{eq:sspal}) is valid, 
and it is possible with some fine-tuning 
to realize $s_{SUSY}$, $t_{SUSY} > 1$.

\section{SUSY saddle point}
\label{sec:susy}

\subsection{General mass formula at SUSY point}
In order to investigate that the SUSY point (\ref{eq:sspal}) 
is stable or not, we examine the mass matrix of $s$ and $t$ 
at this point. It is useful to show general results for 
this matrix, and then we first summarize several formulae 
related to the mass matrix at the SUSY point. 

In general, the first derivative of the scalar 
potential (\ref{eq:scalarpotential}) is
\begin{eqnarray}
\partial_{I} V 
&=&e^{G} \Bigr[
G_I \left( 
G_K G_{\bar{L}} G^{K \bar{L}}-2 \right) 
+G_{\bar{L}} \left( G_{KI}G^{K \bar{L}} 
+ G_K \left(\partial_I G^{K \bar{L}} \right) \right)
\Bigr]. 
\nonumber
\end{eqnarray}
At the SUSY point where $G_I=W^{-1}D_{I}W=0$, 
the second derivatives are shown to be 
\begin{eqnarray}
V_{IJ}&=&-e^{G}G_{IJ}, 
\nonumber \\
V_{\bar{I}\bar{J}}&=&-e^{G}G_{\bar{I}\bar{J}}, 
\nonumber \\
V_{I\bar{J}}&=&e^{G}\left[
-2G_{I\bar{J}}+G_{IK}G_{\bar{J}\bar{L}}G^{K\bar{L}}
\right],
\label{massmt}
\end{eqnarray}
and Hessian matrices of real parts of complex fields 
$I,J$ are given by
\begin{eqnarray}
V_{I_R  J_R} &\equiv& 
V_{IJ}+V_{\bar{I}\bar{J}}+V_{I\bar{J}}+V_{J\bar{I}}, 
\label{realmass}
\end{eqnarray}
where we denoted real parts of $I,J$ as $I_R,J_R$. 
When the K\"ahler potential is a function of only the real part 
of complex fields, $K=K(\phi^I+\bar{\phi}^{\bar{I}})$, and the 
derivatives of superpotential satisfy 
$W^{-1}W_I, W^{-1}W_{IJ} \in {\mathbf R}$, 
we generically have 
\begin{eqnarray}
G_{IJ}=
K_{IJ}-\frac{W_I W_J}{W^2}+\frac{W_{IJ}}{W}=
\bar{G}_{\bar{I}\bar{J}}~\in {\mathbf R}.
\label{cond1}
\end{eqnarray} 

First, for the simplest example, we consider 
a scalar potential of a single complex scalar field $X$. 
In the stationary point of a generic potential, 
we can express mass terms as 
\begin{eqnarray}
V^{mass}&=& V_{X\bar{X}}|x|^2+
\frac{1}{2}\left(~
V_{XX}x^2 +  V_{\bar{X} \, \bar{X}}\bar{x}^2
\right)
\nonumber \\
&=& \left({\rm Re}\,(x),\,{\rm Im}\,(x)
\right)
\left(
\begin{array}{cc}
 V_{X\bar{X}}+{\rm Re}\,(V_{XX}) & -{\rm Im}\,(V_{XX})  \\
 -{\rm Im}\,(V_{XX}) &  V_{X\bar{X}}-{\rm Re}\,(V_{XX})  
\end{array}
\right)
\left(
\begin{array}{cc}
 {\rm Re}\,(x) \\
 {\rm Im}\,(x)  
\end{array}
\right)
\nonumber \\
&=& \left({\rm Re}\,(x'),{\rm Im}\,(x')
\right)
\left(
\begin{array}{cc}
 V_{X\bar{X}}+|V_{XX}| & 0  \\
 0 &  V_{X\bar{X}}-|V_{XX}|  
\end{array}
\right)
\left(
\begin{array}{cc}
 {\rm Re}\,(x') \\
 {\rm Im}\,(x')  
\end{array}
\right), 
\nonumber
\end{eqnarray}
where $x=X-X_0$ and $\partial_X V|_{X=X_0}=0$. 
In the last line we have diagonalized the mass matrix, 
and find that the following condition 
\begin{eqnarray}
V_{X\bar{X}} >|V_{XX}|, 
\label{min}
\end{eqnarray}
is necessary for a point $X=X_0$ to be a minimum of this potential. 
In terms of $G_{IJ}$, $G_{\bar{I}\bar{J}}$ and $G_{I\bar{J}}$, 
the condition (\ref{min}) becomes 
\begin{eqnarray}
|G_{XX}|>2G_{X\bar{X}}.
\nonumber
\end{eqnarray}

Next we consider the case with two moduli fields such 
as $S$ and $T$. We further assume that the K\"ahler 
potential is in a separable form, 
\begin{eqnarray}
K = K^{(S)}(S+\bar{S})+ K^{(T)}(T+\bar{T}),
\nonumber
\end{eqnarray}
so that the K\"ahler metric is diagonal, 
$G_{I\bar{J}}=K_{I\bar{J}}=K_{I\bar{I}}\delta_{I\bar{J}}$. 
At the SUSY point, the mass matrix of the real part fields 
$s={\rm Re}\,S$ and $t={\rm Re}\,T$ 
in the canonical base is expressed by 
\begin{eqnarray}
{\mathcal M}^2 \equiv 
\left(
\begin{array}{cc}
\displaystyle 
\frac{1}{K_{S\bar{S}}} V_{s s} & 
\displaystyle 
\frac{1}{\sqrt{K_{S\bar{S}} K_{T\bar{T}}}} V_{s t}  \\*[15pt]
\displaystyle 
\frac{1}{\sqrt{K_{S\bar{S}} K_{T\bar{T}}}} V_{s t} &  
\displaystyle 
\frac{1}{K_{T\bar{T}}} V_{t t} 
\end{array}
\right), 
\label{eq:massmatrix}
\end{eqnarray}
where each component is given by Eq.~(\ref{realmass}) 
under the reality condition (\ref{cond1}), 
and then we find 
\begin{eqnarray}
{\rm tr}\,{\mathcal M}^2 &=& 
\frac{2e^{G}}{G_{S\bar{S}}G_{T\bar{T}}}
\Big( \frac{G_{T\bar{T}}}{G_{S\bar{S}}} G_{SS}^2 
+\frac{G_{S\bar{S}}}{G_{T\bar{T}}} G_{TT}^2 
+2G_{ST}^2 
\nonumber \\ 
&& \qquad\qquad\qquad 
-4G_{S\bar{S}}G_{T\bar{T}}
-G_{S\bar{S}}G_{TT} 
-G_{T\bar{T}}G_{SS} \Big), 
\nonumber \\
{\rm det}\,{\mathcal M}^2 &=& 
\frac{4e^{2G}}{G_{S\bar{S}}^2G_{T\bar{T}}^2}
\left\{ 
G_{ST}^2-(G_{SS}-2G_{S\bar{S}})(G_{TT}-2G_{T\bar{T}}) \right\} 
\nonumber \\ 
&& \qquad\qquad\qquad \times 
\left\{ 
G_{ST}^2-(G_{SS}+G_{S\bar{S}})(G_{TT}+G_{T\bar{T}})
\right\}. 
\nonumber
\end{eqnarray}
All stable SUSY minima in this case should satisfy 
${\rm tr}\,{\mathcal M}^2 > 0$ and 
${\rm det}\,{\mathcal M}^2 > 0$. 
If either $|G_{SS}| \gg G_{S\bar{S}}$ or 
$|G_{TT}| \gg G_{T\bar{T}}$, 
we have ${\rm tr}\,{\mathcal M}^2 > 0$.
Furthermore, if both $|G_{SS}| \gg G_{S\bar{S}}$ and 
$|G_{TT}| \gg G_{T\bar{T}}$, we have ${\rm det}\,{\mathcal M}^2 > 0$.
However, we obtain ${\rm det}\,{\mathcal M}^2 < 0$, e.g. if 
$|G_{SS}| \gg G_{S\bar{S}}$, 
$|G_{TT}| \sim G_{T\bar{T}}$ and 
$|G_{SS}G_{TT}| \gg G_{ST}^2$.

\subsection{Stability of moduli at SUSY point}
Now we analyze the stability of SUSY stationary point 
in our model. The mass matrix of the fluctuations of 
moduli ($s$,$t$) around the SUSY point is given by 
Eq.~(\ref{eq:massmatrix}). For the parameters satisfying 
Eq.~(\ref{eq:pararef}), we can estimate the mass matrix 
elements by using the approximation (\ref{eq:sspal}) 
and find 
\begin{eqnarray}
{\rm tr}\,{\mathcal M}^2 &\simeq& 
\frac{1}{4} \left| {\rm det}\,{\mathcal M}^2 \right| 
(m_{3/2}^{SUSY})^{-2} 
\ > \ 0, 
\nonumber \\
{\rm det}\,{\mathcal M}^2 &\simeq& 
-\frac{128}{n_S^2} 
\Big( \ln \frac{bB}{aA} \Big)^4 
\frac{(ab)^2}{(b-a)^4} 
(m_{3/2}^{SUSY})^4 
\ < \ 0. 
\nonumber
\end{eqnarray}
Then it is the case that the SUSY stationary point 
is a saddle point and unstable. Actually in the next section 
we will find another stationary point near the SUSY point, 
which is a SUSY breaking (AdS) local minimum. 
The point here is $|G_{SS}| \gg G_{S\bar{S}}$ and that 
the superpotential contributions in $G_{TT}$ are cancelled by 
$G_{ST}^2$ terms in ${\rm det}\,{\cal M}^2$ due to the 
assumption $|b-a| \ll |a|, |b|$ in Eq.~(\ref{eq:pararef}). 
That leads to ${\rm det}\,{\cal M}^2<0$. 

Finally we comment on a {\it trench} of the 
scalar potential along $F^S$- ($F^T$-) flat direction 
in the $(s,t)$ moduli space. For a realistic moduli value, 
$as,bs \gg 1$, a global SUSY part in the scalar potential 
$\left|W_I/W \right|^2$ is dominated for this racetrack model, 
and it is important whether $\alpha, \beta$ is larger or 
smaller than $1$. For $|a|, |b| \gg 1 > |\alpha|, |\beta|$ 
(which is satisfied by Eq.~(\ref{eq:pararef})), the inequality 
$|W_S| \gg |W_T|$ is satisfied and the curve $D_SW \approx 0$ 
determines the structure of the local minimum in the following sense. 
Along this curve, we obtain $\Phi_\bot(s,t) \approx 0$, 
where $\Phi_\bot$ is defined in Eq.~(\ref{eq:massivemode}). 
Actually this $\Phi_\bot$-direction (approximately 
$s$-direction for $|a|, |b| \gg 1 > |\alpha|, |\beta|$) is stabilized 
with a relatively large mass (with canonically 
normalized kinetic term), 
\begin{eqnarray}
m_\bot^2 &\sim& {\rm tr}\,{\mathcal M}^2 
\ \gg \ |m_\parallel^2|,  
\label{eq:massivemodemass}
\end{eqnarray}
due to Eq.~(\ref{eq:pararef}), where $m_\parallel^2$ is the 
mass of the fluctuation $\Phi_\parallel$ along $D_SW \approx 0$ 
perpendicular to $\Phi_\bot$-direction satisfying 
\begin{eqnarray}
|m_\parallel^2| &\sim& 
\left| \frac{{\rm det}\,{\mathcal M}^2}{
{\rm tr}\,{\mathcal M}^2} \right| 
\ \approx \ 4 (m_{3/2}^{SUSY})^2, 
\nonumber
\end{eqnarray}
and $(m_\bot^2,\,m_\parallel^2)$ are the eigenvalues 
of the mass matrix ${\mathcal M}^2$. 
The hierarchy in Eq.~(\ref{eq:massivemodemass}) illustrates that 
the potential of our model has a sharp and deep {\it trench} along 
$D_SW \approx 0$, perpendicular to this $\Phi_\bot$-direction. 
(For $|a|, |b| \gg |\alpha|, |\beta| >1$ or 
$|a\alpha|, |b\beta| \gg 1 > |a|, |b|$, 
the curve $D_TW \approx 0$ determines the local minimum, similarly. 
However the former hardly satisfies $s_{SUSY},t_{SUSY} >1$. 
The latter is just the replacement of $S$ and $T$ essentially.) 
Note that along the {\it trench}, $|G_{IJ}| > 2G_{I\bar{J}}$ 
holds for $^\forall I,J$. Then the saddle point condition, 
${\rm det}\,{\mathcal M}^2<0$, can be simplified as 
\begin{eqnarray}
2\left|{\rm det}\,G \right| &<& 
\left(\,
3+{\rm sign}(G_{IJ})\,{\rm sign}({\rm det}\,G)\,\right)\,
\left(\,G_{S\bar{S}}|G_{TT}|+G_{T\bar{T}}|G_{SS}|\,\right), 
\nonumber
\end{eqnarray}
where ${\rm det}\,G=G_{SS}G_{TT}-G_{ST}^2$, 
under the assumption that all $G_{IJ}$ have the same sign. 

\begin{figure}[t]
\begin{center}
\begin{minipage}{0.48\linewidth}
\epsfig{figure=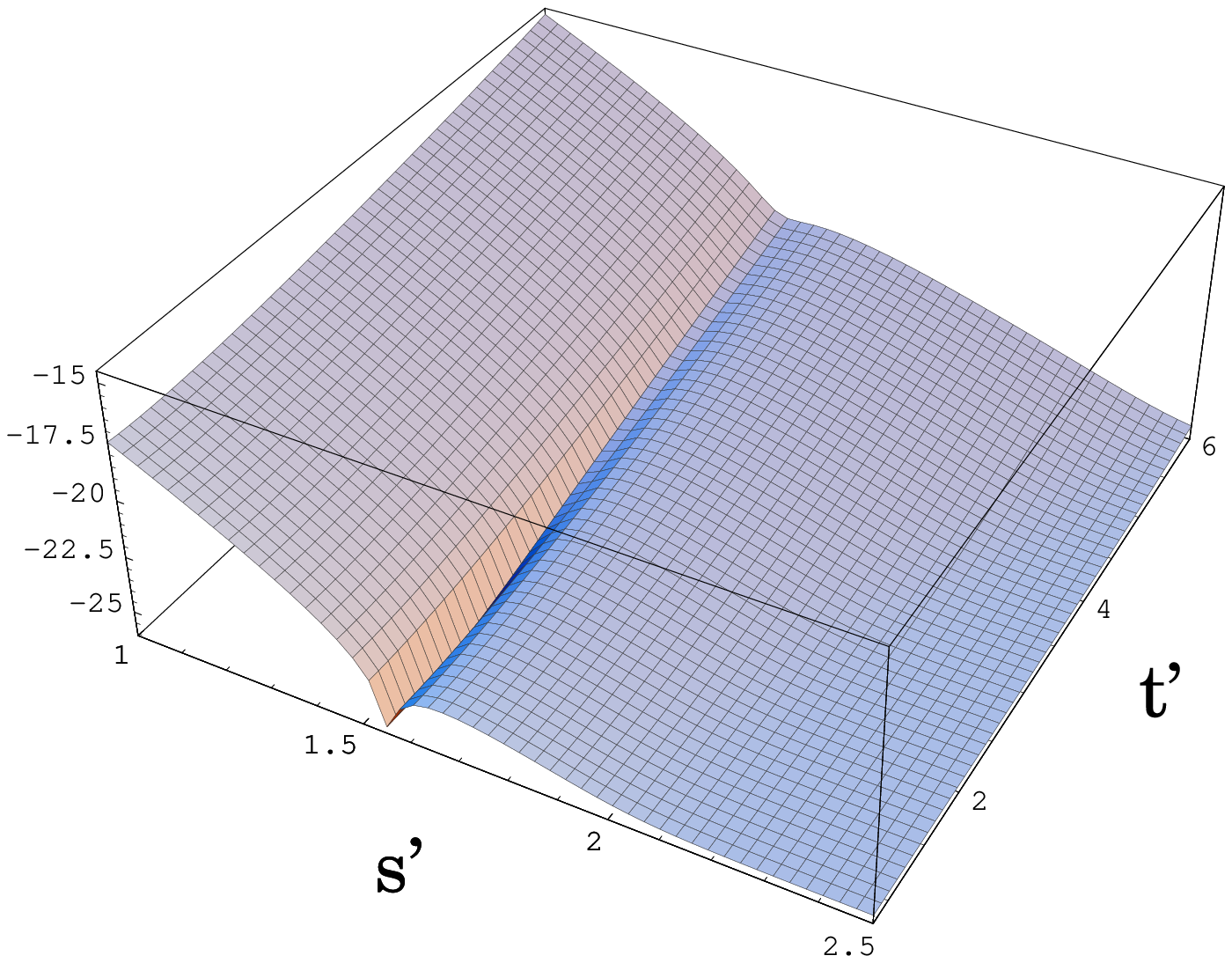,width=\linewidth}
\centerline{(a)}
\end{minipage}
\hfill
\begin{minipage}{0.48\linewidth}
\epsfig{figure=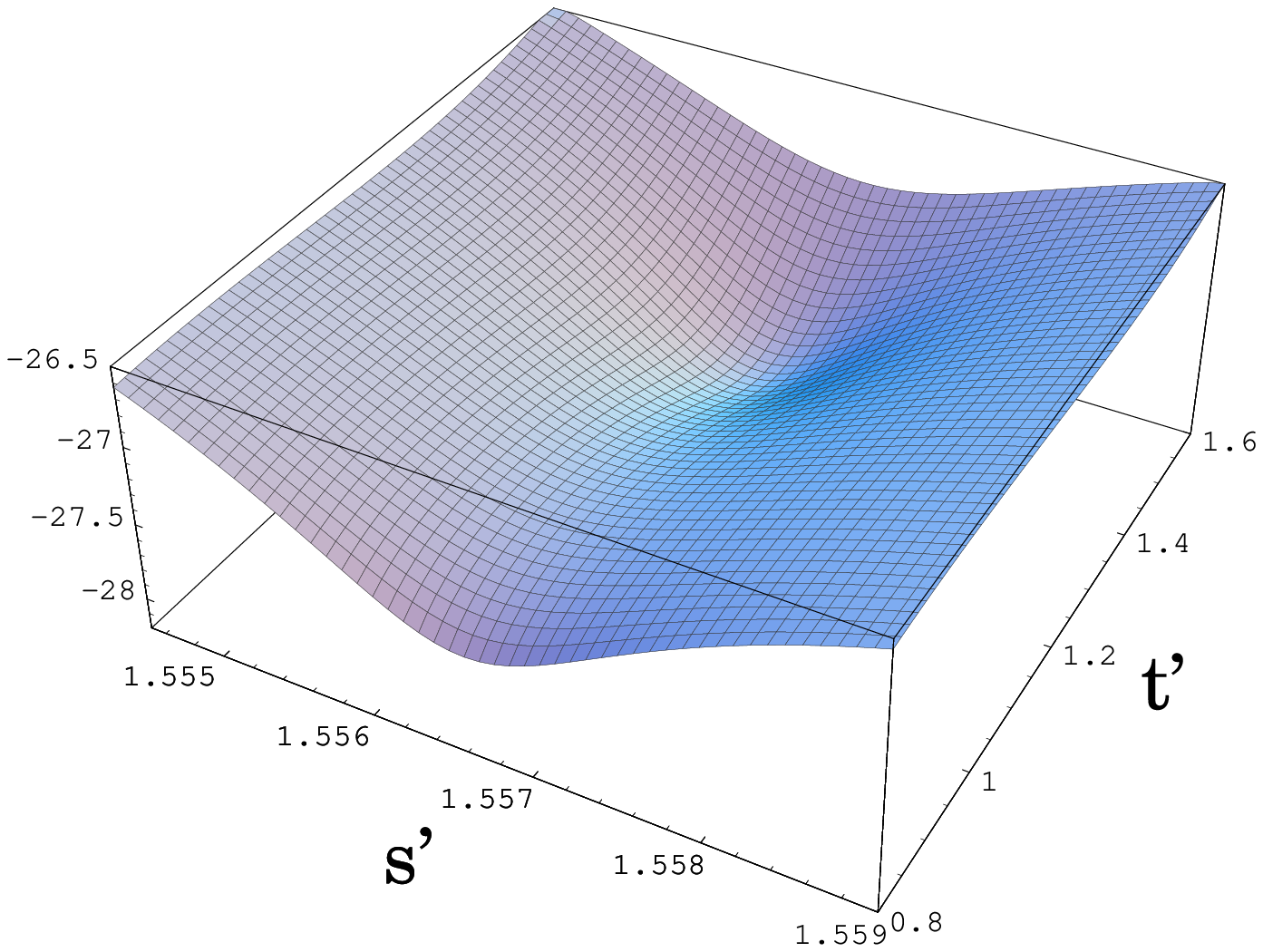,width=\linewidth}
\centerline{(b)}
\end{minipage}
\end{center}
\caption{The logarithm of scalar potential, 
$\log_{10} \big( V-(1+\epsilon)V_{SB} \big)$, 
plotted in $(s',t')$-plane 
where 
$s'=s\,\cos \vartheta-t\,\sin \vartheta$, 
$t'=s\,\sin \vartheta+t\,\cos \vartheta$, 
$\tan \vartheta=-(a\alpha+b\beta)/(b-a)$ 
and $\epsilon=0.01$. 
The parameters are chosen as in Eq.~(\ref{eq:numpara}). 
The sharp and deep {\it trench} along $D_SW=0$ 
curve is shown in (a). 
The SUSY saddle point and the SUSY breaking 
local minimum is shown in (b).}
\label{fig:dsw}
\end{figure}

The structure of the potential along $D_SW=0$ is shown in the 
$(s,t)$-plane in Fig.~\ref{fig:dsw} with the parameter choice 
of Eq.~(\ref{eq:numpara}). In Fig.~\ref{fig:dsw} (a), 
the {\it trench} along $D_SW=0$ is shown clearly, 
and in (b), a region around the SUSY saddle point 
(as well as the SUSY breaking minimum derived in the 
next section) is magnified.

\section{SUSY breaking local minimum}
\label{sec:susybreaking}
To find a true (SUSY breaking) local minimum, we consider 
the case $|a|, |b| \gg |a\alpha|, |b\beta|$ in which there is a sharp 
and deep {\it trench} along the curve determined by $D_S W=0$. 
The transverse mode $\Phi_\bot(s,t)$ to this curve can be 
frozen out as mentioned above, and it is enough to analyze 
the structure of potential along this curve. 

On the curve of $G_S=W^{-1}D_S W=0$ 
with Eq.~(\ref{eq:imaginaryvev}), we generically find 
\begin{eqnarray}
\partial_{t} V \big|_{D_S W=0}(t) 
&=& \big[ 
\big( \partial_t +\partial_{t}{\cal S}(t)\, 
\partial_s \big) V \big]_{D_S W=0}(t) 
\ = \ 
2p(t)\,e^G G_T \big|_{D_S W=0}(t), 
\label{eq:dtv}
\end{eqnarray}
where 
\begin{eqnarray}
p(t) &=& \Bigg[ 
G^{T\bar{T}} \bigg( G_T^2+G_{TT}
+\partial_{t} {\cal S}(t)\, 
G_{ST} \bigg)+G^{T\bar{T}}_{\ \ ,T}G_T -2 
\Bigg]_{D_S W=0}(t), 
\nonumber \\
{\cal S}(t) &=& s\big|_{D_S W=0}(t). 
\nonumber
\end{eqnarray}
Then the stationary points {\it other than} the SUSY point 
($G_T=0$) is determined by $p(t)=0$. 
In the region with $|2a{\cal S}(t)| \gg n_S$, 
the curve $D_S W=0$ is approximated by the asymptotic 
line (\ref{eq:dswal}), 
\begin{eqnarray}
{\cal S}(t) &\simeq& -\frac{a\alpha+b\beta}{b-a}t
+\frac{1}{b-a}\ln \frac{bB}{aA}, 
\label{eq:stal}
\end{eqnarray}
along which $\Phi_\bot(s,t) \simeq 0$ in Eq.~(\ref{eq:massivemode}) 
and then $W_2/W_1 \simeq -a/b$ from Eq.~(\ref{eq:wratio}). 
Due to this, the ratios $W_I/W$, $W_{IJ}/W$ are all 
$t$-independent constants, 
\begin{eqnarray}
&\displaystyle 
W_S/W \ \simeq \ 0, 
\qquad 
W_T/W \ \simeq \ \frac{\alpha+\beta}{b-a}\,ab,& 
\nonumber \\
&\displaystyle 
W_{SS}/W \ \simeq \ -ab, 
\qquad 
W_{TT}/W \ \simeq \ \frac{a\alpha^2-b\beta^2}{b-a}\,ab, 
\qquad 
W_{ST}/W \ \simeq \ -\frac{a\alpha+b\beta}{b-a}\,ab,& 
\nonumber 
\end{eqnarray}
which appear in $G_I=K_I+W^{-1}W_I$ 
and $G_{IJ}=K_{IJ}+W_{IJ}/W-(W_I/W)(W_J/W)$. 
Then the function $p(t)$ defined above is found to be 
\begin{eqnarray}
p(t) &\simeq& 
n_T \Big( \frac{t}{t_{SUSY}} \Big)^2 
-2(n_T-1) \Big( \frac{t}{t_{SUSY}} \Big) 
+n_T-3, 
\nonumber
\end{eqnarray}
where $t_{SUSY}$ is given in Eq.~(\ref{eq:sspal}). 

The SUSY breaking stationary point is determined by 
$p(t_{SB})=0$ that is easily solved as 
\begin{eqnarray}
t_{SB} &\simeq& t_{SUSY}(1+\delta_{SB}^t), 
\label{eq:sbal} \\
s_{SB} &\simeq& {\cal S}(t_{SB}) 
\ = \ s_{SUSY}(1+\delta_{SB}^s), 
\nonumber
\end{eqnarray}
where ($s_{SUSY}$,\,$t_{SUSY}$) is shown 
in Eq.~(\ref{eq:sspal}), 
and\footnote{
Note that the other solution 
$\delta_{SB}^t = -(\sqrt{n_T+1}+1)/n_T$ 
corresponds to 
$t_{SB} \le 0$ for $n_T \le 3$ 
and 
$0< t_{SB} \le t_{SUSY}$ for $n_T > 3$. 
We will focus on the solution (\ref{eq:sbsp}) 
which yields $t_{SUSY}<t_{SB}$, 
although this additional stationary point may 
reside in meaningful region for the latter case 
if $n_T > 3$ is possible.} 
\begin{eqnarray}
\delta_{SB}^t &=& 
\frac{\sqrt{n_T+1}-1}{n_T}, 
\label{eq:sbsp} \\
\delta_{SB}^s &=& 
-\frac{a\alpha+b\beta}{\ln (bB/aA)} \delta_{SB}^t. 
\nonumber
\end{eqnarray}
By noting that 
\begin{eqnarray}
G_T \big|_{D_SW=0}(t) &\simeq& 
\frac{n_T}{2} \Big( \frac{1}{t_{SUSY}}
-\frac{1}{t} \Big), 
\label{eq:gtdsw}
\end{eqnarray}
we can easily examine that this stationary point is a local 
minimum of the scalar potential, due to the fact that 
$e^G>0$, 
$G_T|_{D_SW=0}(t_{SB}) \simeq n_T \delta_{SB}^t/2t_{SB}>0$ 
and 
$p(t_{SB} \pm \epsilon) \gtrless 0$ 
in Eq.~(\ref{eq:dtv}). 
For the parameters satisfying Eq.~(\ref{eq:pararef}), we find 
$\delta_{SB}^t \lesssim 1$ and $\delta_{SB}^s \ \ll \ 1$, 
and then 
$t_{SB} \sim t_{SUSY}>1$, 
$s_{SB} \simeq s_{SUSY}>1$. 

At the SUSY breaking local minimum, the gravitino 
mass $m_{3/2}^2=e^G$ and the order parameter 
$F^T=-K^{T\bar{T}}G_Tm_{3/2}$ are estimated 
respectively as 
\begin{eqnarray}
m_{3/2} &\simeq& 
\frac{e^{-a(\delta_{SB}^s-\alpha \delta_{SB}^t)}}{
(1+\delta_{SB}^s)^{n_S/2}(1+\delta_{SB}^t)^{n_T/2}} 
m_{3/2}^{SUSY}, 
\label{eq:m32sb} \\
\frac{F^T}{T+\bar{T}} &\simeq& -\delta_{SB}^t m_{3/2}, 
\label{eq:ftsb}
\end{eqnarray}
where $m_{3/2}^{SUSY}$ is given in Eq.~(\ref{eq:m32susyads}) 
which can be a TeV scale by tuning parameters, and 
Eq.~(\ref{eq:gtdsw}) has been applied. 
Note that $F^S$ vanishes within the approximation 
along $D_SW=0$ asymptotic line ($|2as| \gg n_S$), 
and it can receive a nonzero contribution 
\begin{eqnarray}
\frac{F^S}{S+\bar{S}} 
&\sim& 
\frac{n_S}{as} \frac{F^T}{T+\bar{T}}. 
\label{eq:gsangle}
\end{eqnarray}
These order parameters generate the vacuum energy 
\begin{eqnarray}
V_{SB} &=& 
n_S \Big| \frac{F^S}{S+\bar{S}} \Big|^2
+n_T \Big| \frac{F^T}{T+\bar{T}} \Big|^2
-3m_{3/2}^2 
\nonumber \\ &\sim& 
\Big( \frac{(\sqrt{n_T+1}-1)^2}{n_T}-3 \Big) m_{3/2}^2 
\ < \ -2 m_{3/2}^2. 
\nonumber
\end{eqnarray}
Therefore we conclude that the local minimum of 
the moduli-mixing racetrack model ($|2as| \gg n_S$) 
provides a SUSY breaking AdS background, which 
generates the moduli-dominated soft SUSY breaking 
terms in the visible sector. 

\begin{table}[t]
\begin{center}
\begin{tabular}{ccccc}
\hline \\*[-10pt]
$s_{SUSY}$ & $t_{SUSY}$ & $V_{SUSY}$ & 
$(m_\bot^{SUSY}/m_{3/2}^{SUSY})^2$ & 
$|m_\parallel^{SUSY}/m_{3/2}^{SUSY}|^2$ \\
\hline \\*[-10pt]
$1.41$ & 
$1.18$ & 
$-1.79 \times 10^{-26}$ & 
$1.11 \times 10^6$ & 
$4.00$ \\
\hline \\*[-10pt]
\hline \\*[-10pt]
$s_{SB}$ & $t_{SB}$ & $V_{SB}$ & 
$(m_\bot^{SB}/m_{3/2})^2$ & 
$(m_\parallel^{SB}/m_{3/2})^2$ \\
\hline \\*[-10pt]
$1.36$ & 
$1.57$ & 
$-1.82 \times 10^{-26}$ & 
$9.55 \times 10^5$ & 
$7.09$ \\
\hline \\*[-10pt]
\hline \\*[-10pt]
$s_{\rm dS}$ & $t_{\rm dS}$ & $V_{\rm dS}$ & 
$(m_\bot^{\rm dS}/m_{3/2}^{\rm dS})^2$ & 
$(m_\parallel^{\rm dS}/m_{3/2}^{\rm dS})^2$ \\
\hline \\*[-10pt]
$1.30$ & 
$1.96$ & 
$+1.20 \times 10^{-34}$ & 
$8.16 \times 10^5$ & 
$4.15 \times 10$ \\
\hline \\*[-10pt]
\hline \\*[-10pt]
AdS & 
$F^S/(S+\bar{S})$ & 
$F^T/(T+\bar{T})$ & 
$m_{3/2}$ & 
$m_{3/2}^{SUSY}$ 
\\*[5pt]
\hline \\*[-10pt]
--- & 
$4.38 \times 10^{-15}$ & 
$-2.77 \times 10^{-14}$ & 
$8.28 \times 10^{-14}$ & 
$7.72 \times 10^{-14}$ \\
\hline \\*[-10pt]
\hline \\*[-10pt]
dS & 
$F^S/(S+\bar{S})$ & 
$F^T/(T+\bar{T})$ & 
$m_{3/2}^{\rm dS}$ & 
$D$ 
\\*[5pt]
\hline \\*[-10pt]
--- & 
$1.34 \times 10^{-14}$ & 
$-6.49 \times 10^{-14}$ & 
$9.78 \times 10^{-14}$ & 
$2.45 \times 10^{-25}$ \\
\hline
\end{tabular}
\end{center}
\caption{A numerical result of the vacuum values 
of moduli fields $(s,t)$, the vacuum energy $V$ and 
the mass eigenvalues $(m_\bot,m_\parallel)$ 
evaluated at the SUSY saddle point (labelled by $SUSY$), 
at the SUSY breaking AdS local minimum (labelled by $SB$) 
and at the uplifted dS minimum (labelled by ${\rm dS}$), 
for the parameter choice of Eq.~(\ref{eq:numpara}) and the 
uplifting potential (\ref{eq:up}) with $(n_P,m_P)=(0,2/3)$. 
The magnitudes of the SUSY breaking order parameters 
at the both AdS and dS minima are also shown. 
The larger (smaller) mass eigenvalue is represented by 
$m_\bot$ ($m_\parallel$) at each point.}
\label{tab:numerical}
\end{table}

A numerical result of the stabilized values of moduli and 
vacuum energy at the SUSY breaking local minimum (as well 
as at the SUSY saddle point shown previously and the 
uplifted local minimum shown later) is shown in 
Table~\ref{tab:numerical} for the parameter choice of 
Eq.~(\ref{eq:numpara}). The SUSY breaking order parameters 
at the minimum are also shown in the table.
The large hierarchy $|m_\bot/m_\parallel| \gg 1$ 
originates from the {\it trench} structure 
of the potential explained in the previous section.

\section{Uplifting}
\label{sec:uplift}
To be phenomenologically viable, we need a Minkowski 
(or dS) vacuum. Unfortunately we could not realize such 
vacuum within our effective 4D SUGRA with two moduli. 
That is rather generic situation.
That is because a SUSY point is a good candidate for 
the potential minimum, but that, in general, leads to 
the negative vacuum energy $V = -3 (m^{SUSY}_{3/2})^2$ 
unless $W=0$ is realized at such SUSY point.\footnote{
In our model, it is impossible to realize $W=0$ at the 
SUSY point with finite moduli values. 
See Appendix~\ref{app:globalsusyvac}.}
Furthermore, if such SUSY point is unstable, 
we would find another SUSY breaking minimum, which has 
of course a negative vacuum energy $V< -3 (m^{SUSY}_{3/2})^2 $.
Thus, the negative vacuum energy is rather generic problem 
within the SUGRA framework.

Here, following the original KKLT scenario~\cite{Kachru:2003aw}, 
we consider a simple deformation of this system by 
introducing additional potential energy, 
\begin{eqnarray}
V_{\rm lift} &=& 
D\,e^{2K/3}(T+\bar{T})^{n_P}(S+\bar{S})^{m_P}, 
\label{eq:up}
\end{eqnarray}
which may arise due to, e.g. the existence of anti $D3$-brane 
at the tip of warped throat of CY space in 
type IIB orientifold models, and breaks 
$N=1$ SUSY explicitly\footnote{
See for possibilities of uplifting in heterotic M-theory, 
e.g. Ref.~\cite{Buchbinder:2004im} and references therein.}. 
We fine-tune the value of $D$ 
in such a way that the vacuum energy of the previous 
local minimum vanishes or becomes slightly positive, 
$$V_{\rm dS} \ = \ V+V_{\rm lift} \ge 0.$$ 

For $s,t \sim 1$, we have $D={\cal O}(m^2_{3/2})$.
However, the mode perpendicular to the deep {\it trench} 
$D_SW=0$  has a larger mass $m^2_\bot \gg m^2_{3/2}$.
Thus, the vacuum shifts little along this direction.
On the other hand, the mode along $D_SW=0$ has a mass 
comparable with the gravitino mass, i.e. 
$m^2_\parallel = {\cal O}(m^2_{3/2})$.
That suggests that the VEV $t_{SB}$ as well as $s_{SB}$ might 
shift by a factor of ${\cal O}(1)$. 
However, the third and higher derivatives of the scalar potential 
at the minimum without uplifting potential are quite large, 
i.e. $\partial^3 V/\partial^2 V = {\cal O}(a)$.
Then, the uplifting potential makes the VEV of $t$ a small 
shift of ${\cal O}(0.1)$. 

For $|2a{\cal S}(t)| \gg n_S$ along $D_SW=0$ 
asymptotic line (\ref{eq:stal}), 
we can estimate the vacuum values of 
$t$ and $s$ with the uplifting (\ref{eq:up}), 
\begin{eqnarray}
t_{\rm dS} &\simeq& t_{SB}(1+\delta_{\rm dS}^t),
\nonumber \\
s_{\rm dS} &\simeq& {\cal S}(t_{\rm dS}) 
\ = \ s_{SB}(1+\delta_{\rm dS}^s),
\nonumber
\end{eqnarray}
at the linear order of $\delta_{\rm dS}^t$. 
By solving the stationary condition 
\begin{eqnarray}
\partial_t \Big( V\Big|_{D_SW=0}(t) 
+V_{\rm lift}\Big|_{D_SW=0}(t) \Big) 
\Big|_{t=t_{\rm dS}} &=& 0, 
\nonumber
\end{eqnarray}
with the fine-tuning condition 
\begin{eqnarray}
D &=& \frac{V_{\rm dS}-V\Big|_{D_SW=0}(t_{\rm dS})}
{(2t_{\rm dS})^{n_P-2n_T/3}(2{\cal S}(t_{\rm dS}))^{m_P-2n_S/3}}, 
\nonumber
\end{eqnarray}
we find, at the linear order of $\delta_{\rm dS}^t$, 
\begin{eqnarray}
\delta_{\rm dS}^t 
&=& \frac{2n_T/3-n_P+(2n_S/3-m_P) 
\partial_t {\cal S}(t_{SB}) (t_{SB}/s_{SB})}
{2n_T/3-n_P+(2n_S/3-m_P) 
\partial_t {\cal S}(t_{SB}) (t_{SB}/s_{SB})^2 
+n_T+1-\sqrt{n_T+1}}, 
\nonumber \\
\delta_{\rm dS}^s 
&=& -\frac{a\alpha+b\beta}{\ln (bB/aA)} 
\delta_{\rm dS}^t, 
\nonumber
\end{eqnarray}
where $\partial_t {\cal S}(t) \simeq -(a\alpha+b\beta)/(b-a)$ 
from Eq.~(\ref{eq:stal}) 
and $V_{\rm dS}/m_{3/2}^2 \ll 1$ has been adopted. 
For $S$-independent uplifting potential $m_P=2n_S/3$, 
the solution is simplified as, 
\begin{eqnarray}
\delta_{\rm dS}^t 
&=& \frac{1}{1+\frac{3(n_T+1-\sqrt{n_T+1})}{2n_T-3n_P}} 
\qquad (m_P=2n_S/3), 
\nonumber
\end{eqnarray}
which is typically ${\cal O}(0.1)$-${\cal O}(1)$ quantity. 

\begin{figure}[t]
\begin{center}
\epsfig{figure=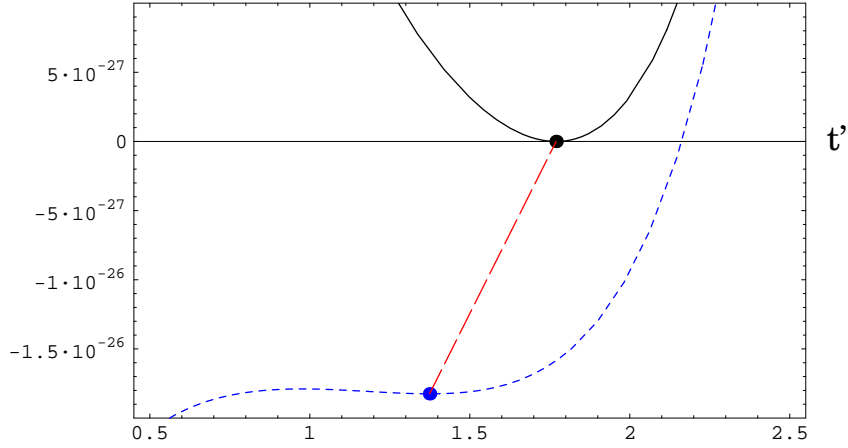,width=0.7\linewidth}
\end{center}
\caption{The behavior of the scalar potential 
$V(s'={s'}_{\rm dS},t')$ including the 
uplifting potential (\ref{eq:up}) with $(n_P,m_P)=(0,2/3)$ and 
the fine-tuned parameter $D=2.45 \times 10^{-25}$ (solid curve), 
as well as $V(s'={s'}_{SB},t')$ without uplifting (dotted curve). 
The parameters are again chosen as in Eq.~(\ref{eq:numpara}). 
The primed-notation 
$(s',t')$ is defined in the caption in Fig.~\ref{fig:dsw} 
and ${s'}_{SB}=s_{SB}\,\cos \vartheta-t_{SB}\,\sin \vartheta$, 
${s'}_{\rm dS}=s_{\rm dS}\,\cos \vartheta
-t_{\rm dS}\,\sin \vartheta$ for the same $\vartheta$. 
The stabilized values $(s_{SB},t_{SB})$ and 
$(s_{\rm dS},t_{\rm dS})$ are shown in 
Table~\ref{tab:numerical}. 
The dashed line represents the shift of the minimum.}
\label{fig:uplift}
\end{figure}

At this dS local minimum with the vacuum energy 
$V_{\rm dS} \ge 0$, 
the gravitino mass 
$(m_{3/2}^{\rm dS})^2=e^G|_{D_SW=0}(t_{\rm dS})$ 
and the order parameter $F^T$ are estimated 
respectively as 
\begin{eqnarray}
m_{3/2}^{\rm dS} &\simeq& 
\frac{e^{-a(\delta_{\rm dS}^s-\alpha \delta_{\rm dS}^t)}}{
(1+\delta_{\rm dS}^s)^{n_S/2}(1+\delta_{\rm dS}^t)^{n_T/2}} m_{3/2}, 
\nonumber \\
\frac{F^T}{T+\bar{T}} &\simeq& 
-(\delta_{SB}^t+\delta_{\rm dS}^t
+\delta_{SB}^t\delta_{\rm dS}^t) m_{3/2}^{\rm dS}, 
\label{eq:ftup}
\end{eqnarray}
where $m_{3/2}$ is shown in Eq.~(\ref{eq:m32sb}), 
and Eq.~(\ref{eq:gtdsw}) has been adopted. 
Again $F^S$ vanishes within the approximation 
along $D_SW=0$ asymptotic line ($|2as| \gg n_S$), 
and it can receive a nonzero contribution 
(\ref{eq:gsangle}) as before. 
Because typically ${\cal O}(\delta_{SB}^t+\delta_{\rm dS}^t
+\delta_{SB}^t\delta_{\rm dS}^t) \sim {\cal O}(\delta_{SB}^t)$, 
we find that the SUSY breaking effect induced by the additional 
explicit breaking term (\ref{eq:up}) does not change the 
qualitative structure of SUSY breaking order parameters from 
the original ones at the local minimum. 
It can change the ratio 
$\frac{F^T}{(T+\bar{T})m_{3/2}}$ 
by at most a factor depending on mainly $n_T$ and $n_P$. 
For example, when $n_S=1$, $n_T=3$, $n_P=0$ and $m_P=2/3$, 
i.e. $V_{\rm lift}=\frac{D}{(T+\bar{T})^2}$, 
the ratio $\frac{F^T}{(T+\bar{T})m_{3/2}}$ becomes 
three times as large as the one in Eq.~(\ref{eq:ftsb}) 
without uplifting. 
Then in this dS (Minkowski) minimum, the relation 
between order parameters again leads to the 
moduli-dominated SUSY breaking. 

Generic form of uplifting potential is not clear still.
However, the third and higher derivatives of our scalar potential 
are quite large as said above.
Thus, when the second and higher derivatives of uplifting potential 
are not large, AdS SUSY breaking vacuum before uplifting 
does not shift drastically.

Table~\ref{tab:numerical} shows a numerical result of 
the stabilized values of moduli and vacuum energy at 
the uplifted dS local minimum by the uplifting potential 
(\ref{eq:up}) with $(n_P,m_P)=(0,2/3)$. 
The parameters are again chosen as in Eq.~(\ref{eq:numpara}), 
and then we fine-tune $D$ as $D=2.45 \times 10^{-25}$ 
in order to realize a dS (almost Minkowski) minimum. 
The SUSY breaking order parameters at the uplifted minimum 
are also shown in the table. The behavior of the total scalar 
potential including the uplifting term with the same 
parameter choice is shown in Fig.~\ref{fig:uplift} 
as well as the original potential without uplifting. 
Note that the stabilized moduli values for more fine-tuned 
$D$ so as to obtain the observed vacuum energy 
$V_{\rm dS} \sim 10^{-120}$ may not be so different 
from the ones shown in Table~\ref{tab:numerical} for 
$V_{\rm dS} \sim 10^{-34}$.

\section{SUSY phenomenology}
\label{sec:SUSYph}

In this section, we discuss SUSY phenomenology of our results.
The size of modulus $F$-term is of ${\cal O}(m_{3/2})$.
Thus, in this type of models, the modulus mediated SUSY 
breaking is dominant.
That is quite different from the original KKLT model 
and modified model with the single light 
modulus \cite{Choi:2004sx,Abe:2005rx},
where anomaly mediation is comparable or rather dominant.

Indeed, dilaton/moduli mediated SUSY breaking has been 
studied \cite{Ibanez:1992hc}.
When there are two and more moduli fields, it is 
phenomenologically useful to 
introduce goldstino angles to parameterize practically 
$F$-terms of moduli without specifying their 
superpotential \cite{Brignole:1993dj,Kobayashi:1994eh,Ibanez:1998rf}.
For example, in our model with two moduli $S$ and $T$, 
we introduce the goldstino angle $\theta$ 
and parameterize moduli $F$-terms,
\begin{equation}
\frac{F^S}{S+\bar S} = C\sqrt{\frac{3}{n_S}} m_{3/2} \sin \theta, \qquad
\frac{F^T}{T +\bar T} =C\sqrt{ \frac{3 }{n_T}} m_{3/2}  \cos \theta,
\end{equation}
up to CP phases.
The $F$-term scalar potential can be written as 
\begin{eqnarray}
V_F &=& \frac{n_S|F^S|^2}{(S+\bar S)^2} + 
\frac{n_T|F^T|^2}{(T+\bar T)^2} - 3 m^2_{3/2}, \\
 &=& 3m^2_{3/2} (C^2 -1) .
\end{eqnarray}
The vanishing $V_F$ corresponds to $C^2=1$, and such 
parameter region has often been used.
However, we would obtain $V_F < 0$ in generic case as 
discussed in the previous section. 
That is, we obtain  $C^2 < 1$, but $C = {\cal O}(1)$. 
For example, we find 
$C \simeq -\sqrt{n_T/3}(\delta_{SB}^t+\delta_{\rm dS}^t
+\delta_{SB}^t\delta_{\rm dS}^t)$ from the result 
in the previous section. 

Our model for 
$|a|, |b| \gg 1 > |\alpha|, |\beta|$ leads to 
$\tan \theta ={\cal O}(1/as) < 1$.
That implies that when the gauge kinetic function of 
the visible sector is also obtained as 
$f_v = S + \gamma T$ with $\gamma < 1$,
the gaugino mass is smaller than the gravitino mass.
Soft scalar masses are naturally of 
${\cal O}(m_{3/2})$ except a particular K\"ahler metric.
On the other hand, when we take 
$|a|, |b| \gg |\alpha|, |\beta| > 1 $, we have 
$\tan \theta > 1$.
Again, the gaugino mass corresponding to 
the gauge kinetic function $f_v = S + \gamma T$ with 
$\gamma > 1$ becomes smaller than the gravitino mass.
Thus, our model provides a concrete model for such 
moduli mediated SUSY breaking.
In our model, whether $\tan \theta $ larger or smaller than unity, 
depends on along which direction the deep {\it trench} is 
$D_S W=0$ or $D_T W=0$.

At any rate, the gaugino masses seem to be suppressed 
compared with the gravitino mass $m_{3/2}$, 
when $\gamma$ corresponds to the parameter region 
similar to $\alpha, \beta$.
As an illustrating example, we use the parameter 
corresponding to Table 1.
In this case, the gaugino mass with the gauge kinetic function 
$f_v = S + \gamma T$ with $\gamma < 0.1$ is obtained as 
\begin{equation}
M_{1/2} = 0.1 \times m_{3/2}.
\end{equation}
On the other hand, the natural order of soft scalar 
masses is of ${\cal O}(m_{3/2})$.

Such spectrum of superpartners would have 
several phenomenological implications.
One of them would realize the focus point \cite{Feng:1999mn}, 
which is important from the viewpoints of 
fine-tuning problem of the minimal supersymmetric 
standard model and dark matter physics.
Such focus point can be realized when 
left- and right-handed squark masses and the 
up-sector Higgs soft mass are degenerate, 
and the gaugino mass as well as the $A$-term 
is smaller.
Such SUSY spectrum can be realized in our model, 
when the K\"ahler metric of left- and 
right-handed quarks and up-sector Higgs fields 
are degenerate.
We need further condition to suppress the A-term.
The SUSY spectrum of our models have other several 
interesting aspects, which would be studied elsewhere.

\section{Conclusions and discussions}
\label{sec:conclusion}
We have investigated supersymmetric models with double 
gaugino condensations in the hidden sector (racetrack model), 
where the gauge couplings depend on two light moduli $S$ and $T$. 
We have analyzed this class of model within the framework 
of effective 4D $N=1$ SUGRA, and have shown that there is 
no stable supersymmetric minimum with finite vacuum values 
of the moduli fields. The true local minimum of the scalar 
potential provides a SUSY breaking AdS background as well as 
reasonable vacuum values of the moduli fields, 
and generate moduli-dominated SUSY breaking soft terms 
for the visible fields. 
This structure of the soft terms may not be affected by the 
uplifting of the minimum by introducing, e.g. anti D3-branes, 
which is required in order to obtain phenomenologically 
viable Minkowski or dS vacuum. During the analysis, we 
have also derived some general formulae related to the 
mass matrix of moduli fields at the SUSY point, which 
would be useful in any case of this kind of analysis. 

Although we have mainly analyzed the case with parameters 
satisfying Eq.~(\ref{eq:pararef}), the resultant local 
structure of the scalar potential around the {\it trench} 
(within the physical domain $s,t>0$) is not changed 
qualitatively even in the other cases, as far as at least one 
of the nonperturbative superpotential terms depends on 
both two moduli, e.g. like the one in the model of 
Ref.~\cite{Becker:2004gw} based on heterotic M-theory with 
open membrane instanton effects, unless there are other effects. 
Then our result may imply that, due to the saddle point structure 
of SUSY stationary point, the anomaly-mediated SUSY breaking 
scenario or the mixed modulus-anomaly 
mediated scenario where 
$m_{3/2}/F^{T,S} \sim 4\pi^2$ are difficult to realize within 
the moduli-mixing racetrack model in which we have two or more 
moduli at low energy, even with the KKLT-type uplifting 
mechanism~\cite{Choi:2004sx}, which could give smaller 
moduli-mediated contributions than the anomaly mediation 
for the single modulus case. 
Note that such small moduli-contribution is possible in 
Refs.~\cite{Choi:2004sx} due to the existence 
of the stable AdS SUSY minimum.

An additional $D$-term potential with pseudo-anomalous $U(1)$
symmetries would change the situation.
Thus, it would be interesting to study the racetrack model 
with the $D$-term corresponding to the pseudo-anomalous $U(1)$
symmetry, where moduli fields transform non-linearly.
However, that is beyond our scope of this paper.

If there exists, e.g. three-form flux in type IIB model 
as in the KKLT model, one of the moduli $S$ can be stabilized 
around the string scale. In this case $S$ should be replaced by 
the vacuum value $\langle S \rangle$ in the effective theory, 
and the phenomenological consequences are quite different 
from those in this paper. 
We have closely studied this possibility 
in Ref.~\cite{Abe:2005rx}.

\subsection*{Acknowledgement}
T.~H.\/ is supported in part by the Grand-in-Aid for Scientific
Research \#171643.
T.~K.\/ is supported in part by the Grand-in-Aid for Scientific
Research \#16028211 and \#17540251.
H.~A.\/, T.~H.\/ and T.~K.\/ are supported in part by 
the Grant-in-Aid for
the 21st Century COE ``The Center for Diversity and
Universality in Physics'' from the Ministry of Education, Culture,
Sports, Science and Technology of Japan.

\appendix
\section{Global SUSY vacuum}
\label{app:globalsusyvac}
In this appendix we show that the global SUSY vacuum 
in our moduli-mixing racetrack model corresponds to 
a runaway solution of the stationary condition. 
The global SUSY vacuum should satisfy $W=0$, 
$W_S=0$ and $W_T=0$ at the same time which yield 
relations between $s$ and $t$ respectively as 
\begin{eqnarray}
t &=& -\frac{b-a}{a\alpha+b\beta}s
+\frac{1}{a\alpha+b\beta}\ln \frac{B}{A}, 
\nonumber \\
t &=& -\frac{b-a}{a\alpha+b\beta}s
+\frac{1}{a\alpha+b\beta} 
\Big( \ln \frac{B}{A}+\ln \frac{b}{a} \Big), 
\nonumber \\
t &=& -\frac{b-a}{a\alpha+b\beta}s
+\frac{1}{a\alpha+b\beta} 
\Big( \ln \frac{B}{A} +\ln \frac{b}{a} 
+\ln \frac{-\beta}{\alpha} \Big). 
\nonumber 
\end{eqnarray}
Only the possibility that these three equations 
can be identical is given by the parameter choice, 
\begin{eqnarray}
a &=& b, 
\qquad 
\alpha \ = \ -\beta, 
\nonumber
\end{eqnarray}
that is we obtain 
\begin{eqnarray}
W &=& (A-B)e^{-a(S-\alpha T)}. 
\nonumber
\end{eqnarray}
However in this case the vacuum with 
$\langle W \rangle=0$ corresponds to 
$a(s-\alpha t) \to \infty$, that is a runaway solution, 
and we can not obtain finite vacuum values of moduli. 
Note that the global SUSY vacuum with finite moduli 
values may be possible if we have a constant piece in 
our superpotential as discussed in, 
e.g. Ref.~\cite{Blanco-Pillado:2005fn}.


\begin{thebibliography}{99}

\bibitem{Affleck:1983mk}
  I.~Affleck, M.~Dine and N.~Seiberg,
  Nucl.\ Phys.\ B {\bf 241}, 493 (1984); 
%
  I.~Affleck, M.~Dine and N.~Seiberg,
  Nucl.\ Phys.\ B {\bf 256}, 557 (1985).

\bibitem{Krasnikov:1987jj}
  N.~V.~Krasnikov,
  %
  Phys.\ Lett.\ B {\bf 193}, 37 (1987); 
%
  T.~R.~Taylor,
  %
  Phys.\ Lett.\ B {\bf 252}, 59 (1990); 
%
  J.~A.~Casas, Z.~Lalak, C.~Munoz and G.~G.~Ross,
  %
  Nucl.\ Phys.\ B {\bf 347}, 243 (1990); 
%
  B.~de Carlos, J.~A.~Casas and C.~Munoz,
  %
  Nucl.\ Phys.\ B {\bf 399}, 623 (1993)
  [hep-th/9204012]; 
%
  M.~Dine and Y.~Shirman,
  %
  Phys.\ Rev.\ D {\bf 63}, 046005 (2001)
  [hep-th/9906246];
%
  T.~Banks and M.~Dine,
  %
  Phys.\ Rev.\ D {\bf 50}, 7454 (1994)
  [hep-th/9406132]; 
%
  P.~Binetruy, M.~K.~Gaillard and Y.~Y.~Wu,
  %
  Nucl.\ Phys.\ B {\bf 481}, 109 (1996)
  [hep-th/9605170]; 
%
  J.~A.~Casas,
  %
  Phys.\ Lett.\ B {\bf 384}, 103 (1996)
  [hep-th/9605180]; 
%
  K.~Choi, H.~B.~Kim and H.~D.~Kim,
  Mod.\ Phys.\ Lett.\ A {\bf 14}, 125 (1999)
  [hep-th/9808122].

\bibitem{Kachru:2003aw}
  S.~Kachru, R.~Kallosh, A.~Linde and S.~P.~Trivedi,
  Phys.\ Rev.\ D {\bf 68}, 046005 (2003)
  [hep-th/0301240].

\bibitem{Giddings:2001yu}
S.~B.~Giddings, S.~Kachru and J.~Polchinski,
Phys.\ Rev.\ D {\bf 66}, 106006 (2002)
[hep-th/0105097].
%
  S.~Kachru, M.~B.~Schulz and S.~Trivedi,
  %
  JHEP {\bf 0310}, 007 (2003)
  [hep-th/0201028]; 
%
  P.~K.~Tripathy and S.~P.~Trivedi,
  %
  JHEP {\bf 0303}, 028 (2003)
  [hep-th/0301139]; 
%
  A.~Giryavets, S.~Kachru, P.~K.~Tripathy and S.~P.~Trivedi,
  %
  JHEP {\bf 0404}, 003 (2004)
  [hep-th/0312104].

\bibitem{Choi:2004sx}
  K.~Choi, A.~Falkowski, H.~P.~Nilles, M.~Olechowski and S.~Pokorski,
  JHEP {\bf 0411}, 076 (2004)
  [hep-th/0411066]; 
%
  K.~Choi, A.~Falkowski, H.~P.~Nilles and M.~Olechowski,
  Nucl.\ Phys.\ B {\bf 718}, 113 (2005)
  [hep-th/0503216].

\bibitem{Choi:2005uz}
  K.~Choi, K.~S.~Jeong and K.~i.~Okumura,
  JHEP {\bf 0509}, 039 (2005)
  [hep-ph/0504037];
%
%
  M.~Endo, M.~Yamaguchi and K.~Yoshioka,
  Phys.\ Rev.\ D {\bf 72}, 015004 (2005)
  [hep-ph/0504036]; 
%
%
A.~Falkowski, O.~Lebedev and Y.~Mambrini,
hep-ph/0507110;
%
%
K.~Choi, K.~S.~Jeong, T.~Kobayashi and K.~i.~Okumura,
hep-ph/0508029. 



\bibitem{Choi:1985bz}
K.~Choi and J.~E.~Kim,
Phys.\ Lett.\ B {\bf 165}, 71 (1985).

\bibitem{Ibanez:1986xy}
L.~E.~Ibanez and H.~P.~Nilles,
Phys.\ Lett.\ B {\bf 169}, 354 (1986); 
%
J.~P.~Derendinger, L.~E.~Ibanez and H.~P.~Nilles,
Nucl.\ Phys.\ B {\bf 267}, 365 (1986); 
%
L.~J.~Dixon, V.~Kaplunovsky and J.~Louis,
Nucl.\ Phys.\ B {\bf 355}, 649 (1991).

\bibitem{Banks:1996ss}
T.~Banks and M.~Dine,
Nucl.\ Phys.\ B {\bf 479}, 173 (1996)
[hep-th/9605136].

\bibitem{Choi:1997an}
K.~Choi,
Phys.\ Rev.\ D {\bf 56}, 6588 (1997)
[hep-th/9706171].

\bibitem{Nilles:1997vk}
H.~P.~Nilles and S.~Stieberger,
Nucl.\ Phys.\ B {\bf 499}, 3 (1997)
[hep-th/9702110].

\bibitem{Buchbinder:2003pi}
E.~I.~Buchbinder and B.~A.~Ovrut,
Phys.\ Rev.\ D {\bf 69}, 086010 (2004)
[hep-th/0310112]; 
%
  A.~Lukas, B.~A.~Ovrut and D.~Waldram,
  Nucl.\ Phys.\ B {\bf 532}, 43 (1998)
  [hep-th/9710208]; 
%
  A.~Lukas, B.~A.~Ovrut and D.~Waldram,
  Phys.\ Rev.\ D {\bf 57}, 7529 (1998)
  [hep-th/9711197].

\bibitem{Cremades:2002te}
D.~Cremades, L.~E.~Ibanez and F.~Marchesano,
JHEP {\bf 0207}, 009 (2002)
[hep-th/0201205].

\bibitem{Lust:2004cx}
D.~Lust, P.~Mayr, R.~Richter and S.~Stieberger,
Nucl.\ Phys.\ B {\bf 696}, 205 (2004)
[hep-th/0404134].


\bibitem{Berg:2004ek}
M.~Berg, M.~Haack and B.~Kors,
Phys.\ Rev.\ D {\bf 71}, 026005 (2005)
[hep-th/0404087].


\bibitem{Abe:2005rx}
  H.~Abe, T.~Higaki and T.~Kobayashi,
  hep-th/0511160.

\bibitem{Ibanez:1992hc}
L.~E.~Ibanez and D.~Lust,
Nucl.\ Phys.\ B {\bf 382}, 305 (1992)
[hep-th/9202046];
%
B.~de Carlos, J.~A.~Casas and C.~Munoz,
Phys.\ Lett.\ B {\bf 299}, 234 (1993)
[hep-ph/9211266];
%
  V.~S.~Kaplunovsky and J.~Louis,
  %
  Phys.\ Lett.\ B {\bf 306}, 269 (1993)
  [hep-th/9303040]. 


\bibitem{Brignole:1993dj}
  A.~Brignole, L.~E.~Ibanez and C.~Munoz,
  %
  Nucl.\ Phys.\ B {\bf 422}, 125 (1994)
  [Erratum-ibid.\ B {\bf 436}, 747 (1995)]
  [hep-ph/9308271].


\bibitem{Kobayashi:1994eh}
T.~Kobayashi, D.~Suematsu, K.~Yamada and Y.~Yamagishi,
Phys.\ Lett.\ B {\bf 348}, 402 (1995)
[hep-ph/9408322]; 
%
A.~Brignole, L.~E.~Ibanez, C.~Munoz and C.~Scheich,
Z.\ Phys.\ C {\bf 74}, 157 (1997)
[hep-ph/9508258].

\bibitem{Ibanez:1998rf}
L.~E.~Ibanez, C.~Munoz and S.~Rigolin,
Nucl.\ Phys.\ B {\bf 553}, 43 (1999)
[hep-ph/9812397].




\bibitem{Randall:1998uk}
  L.~Randall and R.~Sundrum,
  Nucl.\ Phys.\ B {\bf 557}, 79 (1999)
  [hep-th/9810155]; 
%
  G.~F.~Giudice, M.~A.~Luty, H.~Murayama and R.~Rattazzi,
  JHEP {\bf 9812}, 027 (1998)
  [hep-ph/9810442].



\bibitem{Marchesano:2004xz}
  F.~Marchesano and G.~Shiu,
  JHEP {\bf 0411}, 041 (2004)
  [hep-th/0409132].

\bibitem{Cascales:2003zp}
  J.~F.~G.~Cascales and A.~M.~Uranga,
  JHEP {\bf 0305}, 011 (2003)
  [hep-th/0303024].

\bibitem{Aldazabal:1998mr}
  G.~Aldazabal, A.~Font, L.~E.~Ibanez and G.~Violero,
  Nucl.\ Phys.\ B {\bf 536}, 29 (1998)
  [hep-th/9804026].

\bibitem{Abel:2000tf}
  S.~A.~Abel and G.~Servant,
  Nucl.\ Phys.\ B {\bf 597}, 3 (2001)
  [hep-th/0009089]; 
%
T.~Higaki and T.~Kobayashi,
Phys.\ Rev.\ D {\bf 68}, 046006 (2003)
[hep-th/0304200].


\bibitem{Barreiro:1999hp}
  T.~Barreiro and B.~de Carlos,
  JHEP {\bf 0003}, 020 (2000)
  [hep-ph/9912387].

\bibitem{Buchbinder:2004im}
E.~I.~Buchbinder,
Phys.\ Rev.\ D {\bf 70}, 066008 (2004)
[hep-th/0406101].


\bibitem{Feng:1999mn}
J.~L.~Feng, K.~T.~Matchev and T.~Moroi,
Phys.\ Rev.\ Lett.\  {\bf 84}, 2322 (2000)
[hep-ph/9908309]; 
Phys.\ Rev.\ D {\bf 61}, 075005 (2000)
[hep-ph/9909334].


\bibitem{Becker:2004gw}
  M.~Becker, G.~Curio and A.~Krause,
  Nucl.\ Phys.\ B {\bf 693}, 223 (2004)
  [hep-th/0403027].


\bibitem{Blanco-Pillado:2005fn}
J.~J.~Blanco-Pillado, R.~Kallosh and A.~Linde,
hep-th/0511042.





\end{thebibliography}
\end{document}